\documentclass[aps,pra,preprint,groupedaddress]{revtex4-1}

\usepackage{graphicx}
\begin{document}

%\preprint{}

\title{A waveguide cavity resonator source of squeezing.}


\author{M. Stefszky, R. Ricken, C. Eigner, V. Quiring, H. Herrmann, C. Silberhorn}
%\thanks{}
\affiliation{Integrated Quantum Optics, Applied Physics, University of Paderborn, Warburger Strasse 100, 33098 Paderborn, Germany}


\date{\today}

\begin{abstract}
We present the generation of continuous-wave optical squeezing from a titanium-indiffused lithium niobate waveguide resonator. We directly measure $2.9\pm 0.1$\,dB of single-mode squeezing, which equates to a produced level of  $4.9\pm 0.1$\,dB after accounting for detection losses. This device showcases the current capabilities of this waveguide architecture and precipitates more complicated integrated continuous-wave quantum devices in the continuous-variable regime.
\end{abstract}

%\keywords{}

\maketitle

%\section{}
%\subsection{}
%\subsubsection{}

\section{Introduction}

The utility of quantum states of light is continually increasing as the production, manipulation and detection of these states improves. For example, such states are used in gravitational-wave detectors to improve their sensitivity \cite{GWOQE.NPH,LSC13}, whilst quantum key distribution schemes have been demonstrated over useful distances \cite{Korzh15.NP} and are now sold as readily available commercial products.  Concurrently, progress towards quantum computing and quantum information processing in both the continuous-variable (CV) and discrete-variable (DV) regimes is constantly continuing \cite{Menicucci06.PRL,Yukawa08.PRA,Goban15.NC}.

As these technologies continue to develop, the need for devices that are reproducible, integrable, intrinsically stable, compact, and efficient becomes ever more critical. Integrated devices, boasting many of these properties, have made substantial improvements in recent years \cite{Orieux16.JO,Goban15.NC}. Devices have recently been constructed that combine state generation and manipulation in a single chip \cite{Silverstone15.NP,Krapick13.NJP}. Additionally, waveguide devices have recently been used to create more complicated states such as a two-photon NOON state and photon-triplet states \cite{Kruse15.PRA,Krapick16.OE}.

Until very recently most of the attention in integrated quantum optics has been directed towards discrete variable quantum optics \cite{Silverstone15.NP,Goban15.NC}. However, CV quantum optics more easily achieves deterministic, unconditional operation, and is therefore better suited to, or provides alternative solutions to, many tasks. CV quantum optics has been used, for example, to improve spectroscopy \cite{Polzik92.PRL}, metrology \cite{GWOQE.NPH,LSC13} and sensing \cite{Marino11.JMO,Vahlbruch16.PRL} via quantum enhancement. 

The foundation of CV quantum optics is optical squeezing. Traditionally, the strongest levels of squeezing have been produced in second order processes in bulk resonators \cite{Stefszky12.CQG,Mehmet11.OE,Eberle10.PRL}. In contrast, the vast majority of research has been directed towards single-pass pulsed systems \cite{Eto11.OL,Yoshino07.APL,Eto08.OE}. Pulsed systems are used for the intense peak powers, used to drive the nonlinear interaction, and this also helps to overcome the large losses that waveguide systems have traditionally exhibited. Only recently has a fully fibre integrated squeezed light source and detection scheme been demonstrated \cite{Kaiser16.O}. In other work, efficient and complicated manipulation of continuous-variable squeezing has been demonstrated in a silica-on-silicon chip \cite{Masada15.NP} and generation of twin-beam squeezing has been demonstrated in both silicon nitride \cite{Dutt15.PRA} and whispering gallery mode resonators \cite{Furst11.PRL}.

Resonators have the advantage that they allow one to tailor the trade-off between squeezing bandwidth and magnitude for any particular application. Furthermore, they allow for the production of strong continuous-wave squeezing, the magnitude of which is limited in single-pass configurations due to power limitations \cite{Kaiser16.O,Pysher09.OL}. The downside to resonators is twofold; in most circumstances they require active stabilisation and they are generally highly sensitive to intra-cavity losses. If one overcomes these issues then in theory it should be possible to produce high levels of squeezing from integrated systems.

In this paper we present results from a titanium-indiffused lithium niobate waveguide resonator source of single-mode squeezed light that produces high levels of continuous-wave squeezing, on par with the best squeezing produced in waveguides from any system. The strong levels of squeezing that are generated precipitates the development of more complicated CV integrated quantum optics devices. Further improvements to increase the performance and utility of the device are discussed.

\section{The Device}

Waveguides are fabricated in z-cut LiNbO$_3$ by an indiffusion of lithographically patterned 7 $\mu$m wide, 80 nm thick titanium strips. The diffusion is performed at 1060 $^\circ$C for 9 hours in oxygen atmosphere. In a subsequent second lithography step, an insulating photoresist pattern is defined which is used for field assisted periodic domain inversion. We have chosen a poling period of approximately 16.9 $\mu$m  to achieve the desired Type 0 quasi-phase matching, where all interacting fields are in the TM polarization, at approximately 170 degrees Celsius. This waveguide architecture has been chosen because of the ability to produce extremely low loss waveguides (0.02dB/cm at 1550nm) \cite{Luo15.NJP}. An 80mm sample is produced which is cut into a number of smaller pieces that are used for squeezing, second harmonic production, and local oscillator shaping.

The end-face coating, deposited in-house, which is used as the output of the device has a reflectivity of $77\pm1\%$ at the fundamental wavelength, chosen to optimise the escape efficiency, $\eta_{esc} = 81\pm 1\%$, whilst keeping the threshold power at a reasonable level, $P_{th} \approx 100$\,mW. This is particularly important for this type of waveguide because photorefractive effects will limit the device performance at high pump powers \cite{Fontana01.OM,Carrascosa08.OE}. The other surface has a high-reflectivity (HR) coating at the fundamental and both faces are anti-reflecting (AR) at the second harmonic. By looking at the transmitted and reflected power from the cavity (far from phase matching), both on and off-resonance \cite{Regener85.APB}, we are able to determine the reflectivity of the HR coating (R=99.0$\pm0.1$) and the intra-cavity loss (0.13$\pm0.01$dB/cm), which are consistent with expected values for this sample and from which the escape efficiency and other cavity properties are determined.

The cavity length is stabilised using only a standard temperature feedback scheme, one that does does not sense the cavity resonance condition. A two stage oven is used in which a larger thermal mass is heated to temperatures of around 160 degrees Celsius using a resistive heating cartridge and fast fine-tuning is achieved via a thermoelectric device. The entire setup is then enclosed in boxes resulting in a system that is able to hold cavity resonance after initial thermalisation for time scales up to hours, which equates to a temperature stability on the millikelvin scale. This high operating temperature is chosen in order to reduce the effect of photorefraction.

\section{Experimental Setup}

\begin{figure}
  \includegraphics[width=0.9\linewidth]{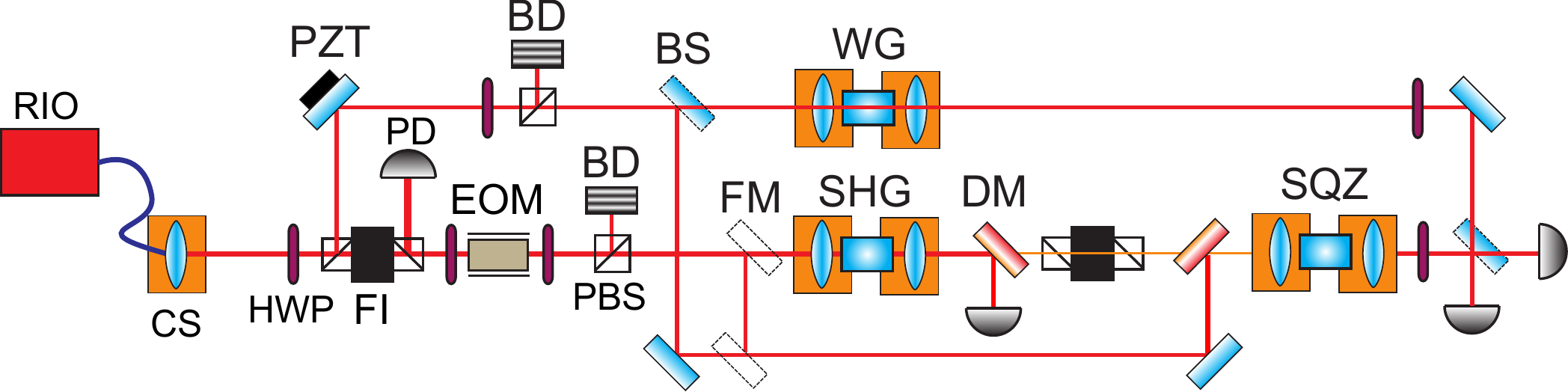}
  \caption{Experimental schematic of the squeezer setup. RIO = RIO Grande laser system, CS = coupling stage, HWP = half-wave plate, PD = photodetector, BD = beam dump, PZT = Piezoelectric transducer, FI = Faraday isolator, EOM = electro-optic modulator, PBS = polarising beam splitter, BS = beam splitter, FM = flipper mirror, WG = waveguide, SHG = second harmonic waveguide resonator, SQZ = squeezer waveguide resonator.}
  \label{experiment}
\end{figure}

The experimental layout is shown in Figure \ref{experiment}. A 1550\,nm RIO Grande laser system provides up to 1 Watt of laser power. The power can be distributed between the local oscillator arm and the second harmonic stage using a Faraday isolator and half-wave plate combination. The light that passes through the Faraday isolator is frequency modulated at 25 MHz before entering the SHG cavity. SHG is achieved in a waveguide resonator that is essentially a copy of the squeezer. Up to 40\,mW of second harmonic power is stably produced by the second harmonic cavity with a pump power of approximately 60\,mW. The cavity resonance condition is locked using a Pound-Drever-Hall scheme \cite{Black01.AJP}. The aforementioned frequency modulation passes through the cavity and is detected by photodetector after a dichroic mirror. This  photocurrent is then fed into a Toptica Digilock system wherein an error signal is produced that is fed back to the frequency of the laser. 

The generated second harmonic power is coupled to the squeezer cavity after passing through a number of dichroic mirrors and a Faraday isolator for the second harmonic. This field makes a single-pass of the cavity, producing squeezing in the waveguide cavity mode. Detection of the squeezed field produced is then achieved via standard homodyne detection. The detector itself consists of two Hamamatsu G12180 photodiodes (quantum efficiency of $\approx 88\%$) in a current subtraction setup \cite{Stefszky12.CQG} with fine tuning of the subtraction achieved via a half-wave plate in the local oscillator arm, which varies the splitting ratio on a 50:50 beasmplitter. Common-mode rejection of up to 50\,dB was seen when a 100\,kHz modulation was applied to the laser. Linearity of the measured shot noise level was observed with local oscillator powers from 2\,mW to 16\,mW. Dark noise clearance of approximately 11dB is seen with a 2\,mW local oscillator.

In order to optimise the spatial overlap between the squeezed field and the local oscillator (LO) the LO is transmitted through a waveguide (cut from the same large sample as the squeezer and SHG resonator) whose temperature is allowed to drift at room temperature, far from phase matching. Using this method we were able to achieve a visibility between the local oscillator and the squeezed fields of up to $97\%$. Visibility is measured with the use of a flipper mirror before the SHG cavity, used to direct this power through the on-resonance squeezer, after which it interferes with the local oscillator. The visibility achieved during the presented data run of $95\pm1\%$ combined with the propagation efficiency, $\eta_{p} = 92\pm 2 \%$, and the aforementioned detector efficiency results in an expected total detection efficiency of $73 \pm 3\%$. 

\section{Results and Discussion}

To characterise the performance of the waveguide resonator, we first investigate the nonlinear gain of the device. The gain of the device (assuming that the nonlinear gain is real) can be described by \cite{Lam99.JOB},
\begin{eqnarray}
G = \frac{(1\pm \sqrt{P/P_{th}})^2}{(1-P/P_{th})^2},
\end{eqnarray}
where $G$ is the (below-threshold) parametric gain, defined as the ratio between the fundamental power exiting the cavity with gain and the power exiting the cavity without gain.

\begin{figure}
  \includegraphics[width=0.7\linewidth]{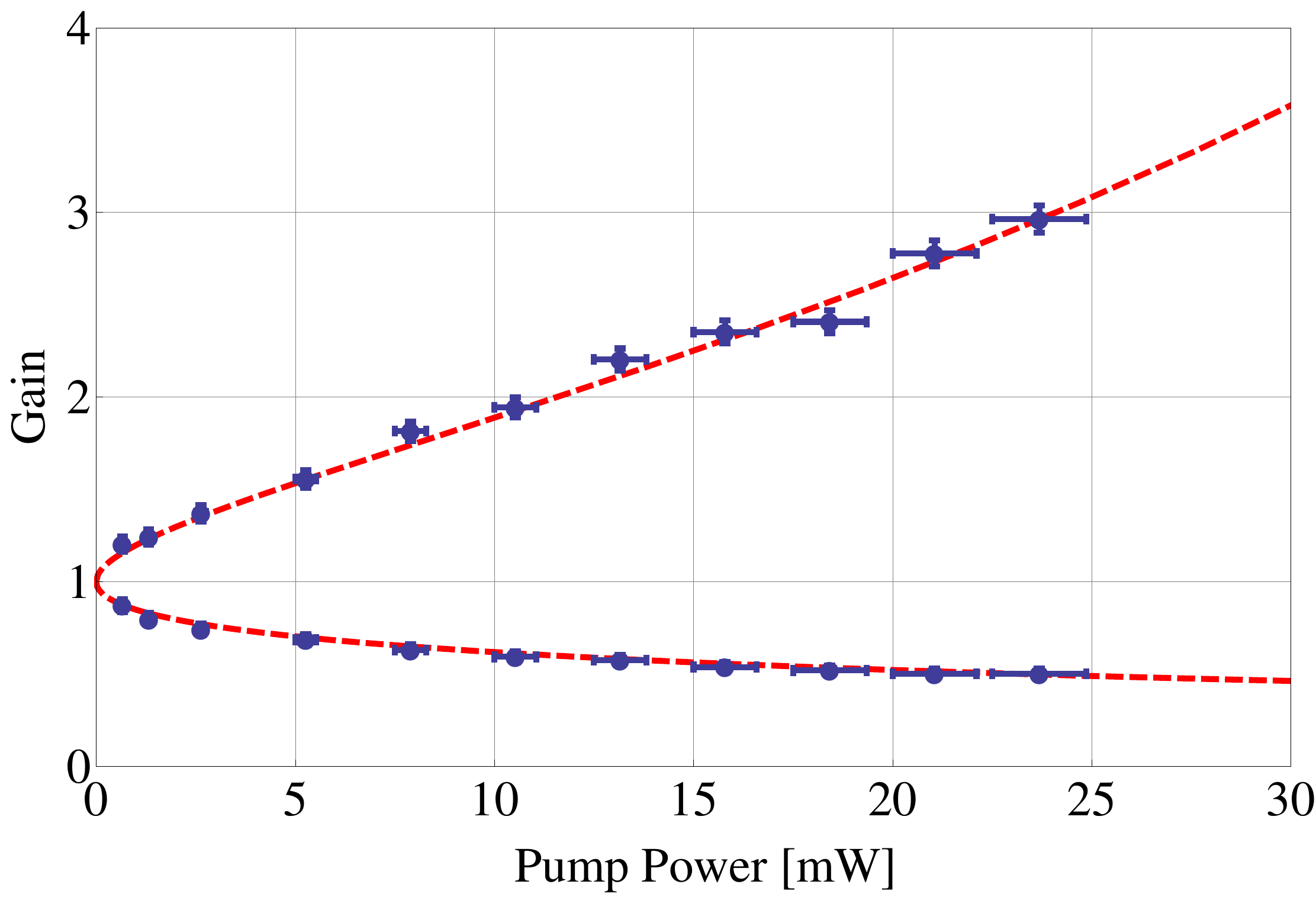}
  \caption{Parametric amplification and de-amplification of a weak seed field as the pump power is varied. The pump power is defined as the power incident on the waveguide. The fit (dashed red line) is made with a threshold power of $135$\,mW. Error bars are shown for data points with uncertainties larger than the marker size.}
  \label{gain}
\end{figure}

A beamsplitter in the LO path (shown in Figure \ref{experiment}) provides a weak seed field that enters the device together with the second harmonic pump. A piezo-electric transducer (PZT) in the seed beam is used to scan the phase between the two fields, such that amplification and deamplification of the seed can be observed. After traversing the dichroic mirrors and Faraday isolator, the 40\,mW SH power produced by the SHG cavity is reduced to approximately 23$\,$mW, measured directly in front of the squeezer cavity. A half-wave plate before the Faraday isolator allows one to vary the amount of power entering the squeezer.  The observed gains as the pump power was varied are shown in Figure \ref{gain}.

We see that the experimental data agrees very well with the theory. The fit indicates that the system has a threshold power (here defined as the power \textit{incident} on the waveguide) of approximately 135\,mW. Unfortunately this number is not an accurate value for the true threshold power of the resonator itself because the waveguide is multi-mode at 775\,nm. It is therefore very difficult to determine the exact amount of power in the fundamental mode, which is the only mode that couples with the squeezed field for the given phase matching conditions.

The seed field is then blocked, leading to the production of vacuum squeezing. We expect the detected noise variances for the squeezed and anti-squeezed quadratures $V_{\pm}$, to follow the standard (non pump-depleted) theory \cite{Lam99.JOB},
\begin{eqnarray}
V_{\pm} &=& 1\mp \eta_{esc} \eta_{det}\frac{4  \sqrt{P/P_{th}}}{(1\pm \sqrt{P/P_{th}})^2},
\end{eqnarray}
where the escape efficiency is given by $\eta_{esc} = \frac{\gamma_{a}^{coup}}{\gamma_{a}^{tot}}$, and $\eta_{det}$ represents the total detection efficiency (propagation, photodetector and homodyne detector efficiencies). The cavity decay rates are defined using $\sum \gamma_a^{i} = (1-\sqrt{R_i})/ \tau$, where the $R_i$ are the equivalent power reflectivities for each loss source and $\tau$ is the round-trip time of the resonator.  The final terms are the pump field power $P$, and the threshold pump field power $P_{th}$. It has been assumed that the sideband measurement-frequency is well within the bandwidth of the cavity and that phase fluctuations can be neglected \cite{Dwyer12.LTN}.

\begin{figure}
  \includegraphics[width=0.7\linewidth]{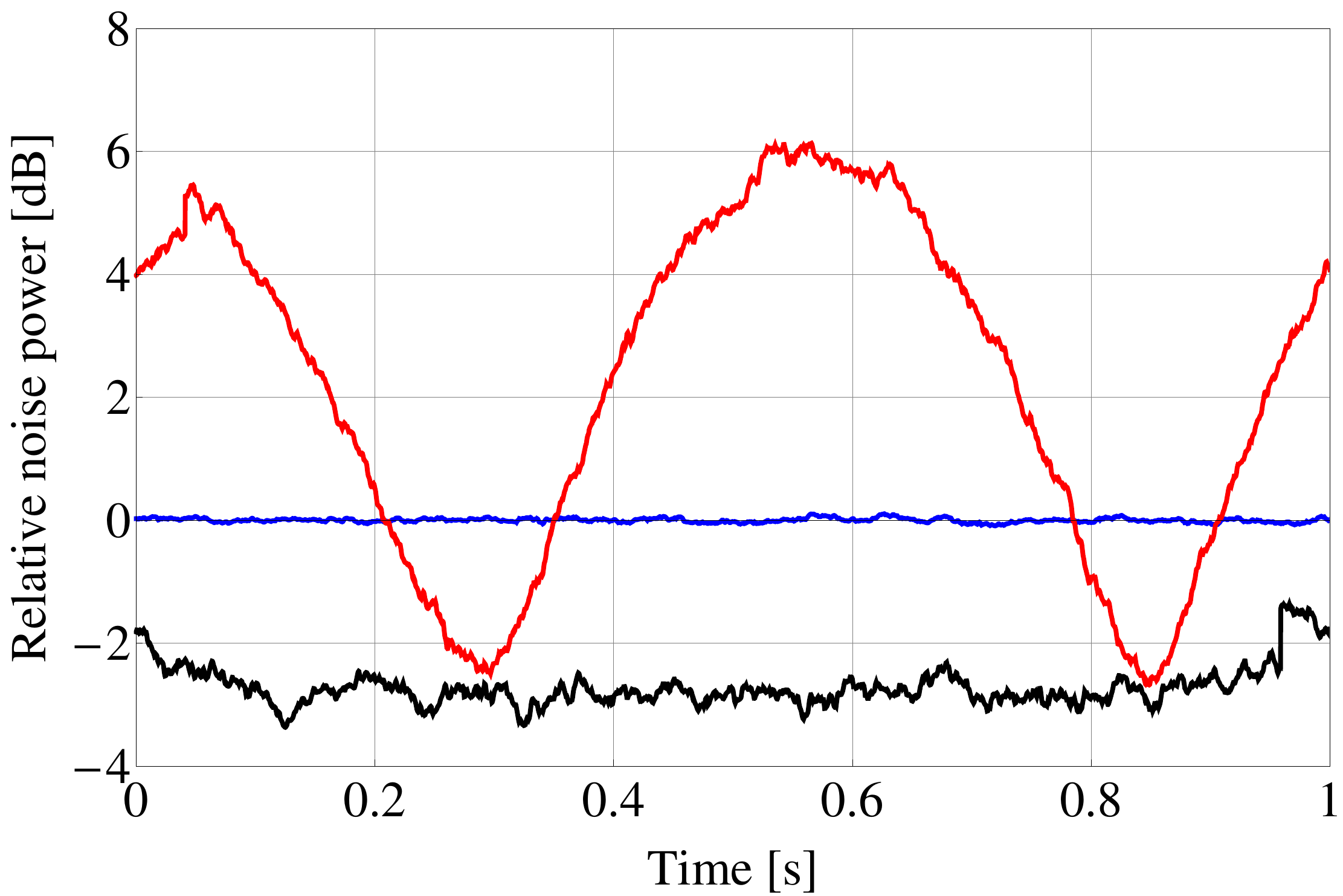}
  \caption{Measured noise variances of the shot noise (blue), scanned squeezing (red), and drifting squeezing (black). The shot noise has been averaged 20 times. RBW = 20kHz, VBW = 10Hz for all traces. Dark noise is approximately 12dB below shot noise. All data has been corrected for dark noise.}
  \label{scan}
\end{figure}

The quadrature measured by the homodyne is varied by scanning the phase of the local oscillator using the same PZT as before. This allows one to measure the squeezing and anti-squeezing. In order to more accurately measure the levels of squeezing, the phase scan is turned off and the phase is allowed to drift. The (dark noise corrected) results, relative to the shot noise (averaged 20 times), are shown in Figure \ref{scan}. With the maximum pump power $2.9\pm 0.1\,$dB of squeezing and $6.0\pm 0.1\,$dB of anti-squeezing is observed. A fit to these parameters shows that these results are consistent with a total state loss of $58\,\%$. Removing the measured escape efficiency from this value reveals a total detection efficiency of $72\,\%$, which is in agreeance with the previously determined value for the detection losses of $73\pm 3\,\%$.

\begin{figure}
  \includegraphics[width=0.7\linewidth]{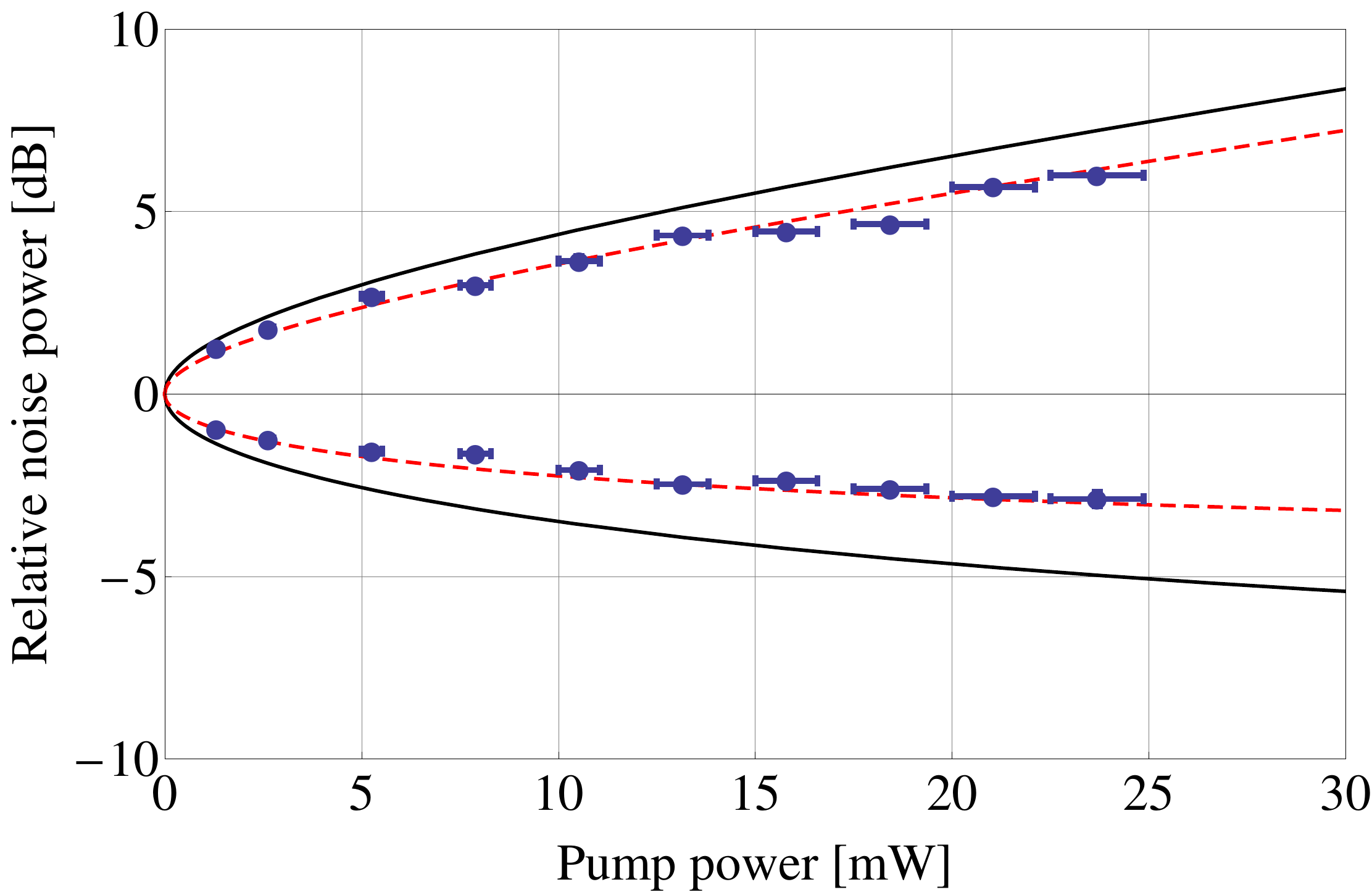}
  \caption{Measured squeezing and anti-squeezing levels as the pump power is varied. The red (dotted) trace is a fit including detection losses and the black (solid) trace shows the squeezing and anti-squeezing levels exiting the device, calculated by correcting for detection losses. Error bars are shown for points where the error is greater than the marker size.}
  \label{squeezing}
\end{figure}

Finally, the previous measurement from Figure \ref{scan} was repeated a number of times with differing pump powers. Each time the power is varied the system is given a few minutes to stabilise. After this thermalisation time, the cavity resonance is stable over times scales of at least a number of minutes, more than enough time to produce reliable squeezing results. From these traces the amount of anti-squeezing and squeezing were recorded, and are plotted in Figure \ref{squeezing} as a function of the pump power \textit{incident on the cavity.}

We immediately see that the measured squeezing and anti-squeezing levels fit the predicted behaviour very well, indicating that the prior assumptions are valid and giving confidence in the previously determined losses. The red dotted trace is a fit to the data with the threshold power of the cavity as the only free parameter. The black trace then illustrates the amount of squeezing produced by the cavity (found by removing the detection losses).

The device can be further improved through a number of methods. Certainly the device can be improved by reducing the losses. Titanium-indiffused lithium niobate waveguides have been shown to have losses as low as 0.02\,dB/cm \cite{Luo15.NJP}. Although some of the waveguides on the current sample had losses below 0.1\,dB/cm, the nonlinear efficiency or phasematching temperature at these waveguides were not desirable. If the losses were to be reduced from 0.13\,dB/cm to 0.02\,dB/cm, then this sample would produce over 8\,dB of squeezing with no other changes to mirror coatings or pump power. Further optimisation and an increase in the available pump power would allow for the production of yet more squeezing. Although increasing the pump power will increase the squeezing, it remains to be seen at what pump power levels photorefraction will begin to affect the properties of the squeezed state.

Another improvement to the device would be to ensure long-term stability through active locking of the resonator length. Exploiting the fact that this waveguide technology has desirable electro-optic properties, a phase modulator could be added to this device to for this purpose. With the correct locking scheme, a phase modulator will allow for fast locking of the waveguide cavity length, even when producing vacuum squeezing \cite{Vahlbruch06.PRL}.

\section{Conclusions and Outlook}

We have presented a titanium-indiffused lithium niobate waveguide resonator that produces up to $4.9\pm 0.1$\,dB of squeezing. From this state we were able to directly measure $2.9\pm 0.1$\,dB of single-mode vacuum squeezing. By increasing pump power, reducing losses and with further resonator optimisation, we expect that it will be possible to produce greater than 8\,dB of squeezing using this architecture. The addition of an on-chip modulator would also allow for a robust, compact solution to ensuring cavity stability over time. The results presented here show that integrated platforms have now progressed to a stage where it is possible to produce useful levels of squeezing without the need for pulses. This opens up the possibility to use these resources in integrated systems to provide quantum enhancement for various applications.

\begin{acknowledgments}
The authors acknowledge funding from the Gottfried Wilhelm Leibniz-Preis.
\end{acknowledgments}

\bibliography{2017MarBib}

\clearpage

\end{document}